\documentclass{article}
\usepackage{ijcai13}
\usepackage{times}

\usepackage{times}
\usepackage{helvet}
\usepackage{courier}
\usepackage{amsmath}
\usepackage{amssymb}
\usepackage{algorithm}
\usepackage[noend]{algorithmic}
\usepackage{color}
\usepackage{graphicx}
\usepackage[chatter]{rotating}
\usepackage{pstricks-add}
\usepackage{pst-tree}
\usepackage{multirow}

\newtheorem{thm}{Theorem}[section]
\newtheorem{cor}[thm]{Corollary}

\newtheorem{defn}[thm]{Definition}

\title{On the complexity of strong Nash equilibrium:\\
Hard--to--solve instances and smoothed complexity}

\author{Nicola Gatti \and Marco Rocco\\
Politecnico di Milano\\
Piazza Leonardo da Vinci 32\\
Milano, Italy\\
\{ngatti, mrocco\}@elet.polimi.it\\
\And
Tuomas Sandholm\\
Carnegie Mellon\\
Computer Science Department\\
5000 Forbes Avenue\\
Pittsburgh, PA 15213, USA\\
sandholm@cs.cmu.edu}

\begin{document}

\maketitle

\begin{abstract}
The computational characterization of game--theoretic solution concepts is a central topic in artificial intelligence, with the aim of developing computationally efficient tools for finding optimal ways to behave in strategic interactions. The central solution concept in game theory is \emph{Nash equilibrium} (NE). However, it fails to capture the possibility that agents can form coalitions (even in the 2--agent case). \emph{Strong Nash equilibrium} (SNE) refines NE to  this setting. It is known that finding an SNE is $\mathcal{NP}$--complete when the number of agents is constant. This hardness is solely due to the existence of mixed--strategy SNEs, given that the problem of enumerating all pure--strategy SNEs is trivially in $\mathcal{P}$. 
Our central result is that, in order for a game to have at least one non--pure--strategy SNE, the agents' payoffs restricted to the agents' supports must, in the case of 2 agents, lie on the same line, and, in the case of $n$ agents, lie on an $(n-1)$--dimensional hyperplane. Leveraging this result, we provide two contributions. First, we develop worst--case instances for support--enumeration algorithms. These instances have only one SNE and the support size can be chosen to be of any size---in particular, arbitrarily large.
Second, we prove that, unlike NE, finding an SNE is in smoothed polynomial time: generic game instances (i.e., all instances except knife--edge cases) have only pure--strategy SNEs.
%
%
\end{abstract}

\section{Introduction}
\vspace{-0.1cm}
The computational characterization of game--theoretic solution concepts is a central topic in artificial intelligence, with the aim of developing computationally efficient tools for finding optimal ways to behave in strategic interactions.  The central solution concept provided by game theory is \emph{Nash equilibrium} (NE). Every finite game admits at least one NE in mixed strategies. Computer scientists have characterized the complexity of NE finding and provided a number of algorithms. Finding an NE of a strategic--form (aka normal-form) game is $\mathcal{PPAD}$--complete~\cite{papastoc} even with just two agents~\cite{Chen09:Settling}. Such games are called bimatrix games.  Although $\mathcal{PPAD} \subseteq \mathcal{NP}$, it is generally believed that $\mathcal{PPAD} \neq \mathcal{P}$ and therefore that there not exists any polynomial--time algorithm to find an NE unless $\mathcal{P}=\mathcal{NP}$. Furthermore, bimatrix games do not have a fully polynomial--time approximation scheme unless $\mathcal{PPAD} \subseteq \mathcal{P}$~\cite{Chen09:Settling} and finding an NE in bimatrix games is not in smoothed--$\mathcal{P}$ unless $\mathcal{PPAD} \subseteq \mathcal{RP}$~\cite{chen} and, therefore, by definition of smoothed complexity, game instances remain hard even if subjected to small perturbations. Instances~\cite{Savani06hard-to-solvebimatrix} that require exponential time when solved with a number of algorithms~\cite{lemke1964,porter2004} are known. (However, these instances are unstable: they become easy when small perturbations are applied~\cite{UAI2012}.)

NE captures the situation in which no agent can gain more by unilaterally changing her strategy.
When agents can form coalitions and change their strategies multilaterally in a coordinated way, the most natural solution concept is \emph{strong Nash equilibrium} (SNE)~\cite{aumann1960}.  An SNE is a strategy profile from which no coalition can deviate in a way that benefits each of the deviators.  
SNE has significantly different properties than NE. An SNE is not assured to exist.  Finding an SNE (determining if one exists) is $\mathcal{NP}$--complete when the number of agents is constant; $\mathcal{NP}$--hardness was proven in~\cite{sandholmComplexiy2008} and membership in $\mathcal{NP}$  in~\cite{aamasSNE2013}.  Unlike for NE, the literature has very few algorithms for SNE.  There are algorithms for finding pure--strategy SNEs for specific classes of games, e.g., congestion games~\cite{holzman1997,HayrapetyanSTOC2006,RozenfeldWINE2008,Hoefer:2010:CPN:1929237.1929264}, connection games~\cite{EpsteinACMEC2007},  maxcut games~\cite{GourvesWINE2009}, and continuous games~\cite{nessah2010}. However, the hardness in the general setting is due to the existence of mixed--strategy SNEs, given that the pure--strategy ones can be found in polynomial time by combining support enumeration and verification.  The only prior SNE--finding algorithm that works also for mixed strategies is only for 2--agent games~\cite{aamasSNE2013}. Its application to instances from the ubiquitous NE benchmark testbed, GAMUT~\cite{gamut}, shows that these instances either admit pure SNEs or do not admit any SNE and therefore that new benchmark testbeds for SNE--finding algorithms are needed.

In this paper, we provide the following main contributions.
\begin{itemize}
\item We show that, if there is a mixed--strategy SNE, then the payoffs restricted to the actions in the support must satisfy specific conditions.  For example, in 2--agent games, they must lie on the same line in agents' utilities space.
\vspace{-0.1cm}
\item We show how to generate 2--agent games with $m$ actions per agent, only one SNE, and any desired number of actions $\{1,\ldots,\frac{m}{2}\}$ in the support of each agent's mixed strategy.  These are the worst--case instances for support--enumeration algorithms, requiring time $O(4^{\frac{m}{2}})$.
\vspace{-0.1cm}
\item We show that, for any number of agents, finding an SNE is in smoothed--$\mathcal{P}$, thus admitting a deterministic support--enumeration algorithm with smoothed polynomial running time.  In the generic case (i.e., in all except knife--edge cases), all SNEs are pure.
\end{itemize}
\vspace{-0.5cm}

\section{Preliminaries}
\vspace{-0.1cm}
A strategic--form game is a tuple $(N, A, U)$ where~\cite{shoham-book}:
\begin{itemize}
\vspace{-0.1cm}
\item $N=\{1,\dots,n\}$ is the set of agents (we denote by~$i$ a generic agent),
\vspace{-0.1cm}
\item $A=\{A_1,\ldots,A_n\}$ is the set of agents' actions and $A_i$ is the set of agent~$i$'s actions (we denote a generic action by $a$,  and by $m_i$ the number of actions in $A_i$),
\vspace{-0.1cm}
\item $U=\{U_1\ldots,U_n\}$ is the set of agents' utility arrays where $U_i(a_1,\ldots,a_n)$ is  agent~$i$'s utility when the agents play actions $a_1,\ldots,a_n$.
\end{itemize}
\vspace{-0.1cm}
We denote by $x_{i}(a_i)$ the probability with which agent~$i$ plays action $a_i\in A_i$ and by $\mathbf{x}_i$ the vector of probabilities $x_i(a_i)$ of agent~$i$. We denote by $\Delta_i$ the space of well--defined probability vectors over $A_i$. We denote by $S_i$ the support of agent~$i$, that is, the set of actions played with positive probability, and by $S$ the support profile $(S_1,\ldots,S_n)$.

The most central solution concept in game theory is NE. A strategy profile $\mathbf{x}=(\mathbf{x}_{1},  \ldots, \mathbf{x}_{n})$ is an NE if, for each $i\in N$, $\mathbf{x}_{i}^{T}U_{i}\prod_{j\neq i}\mathbf{x}_{-j} \geq \mathbf{x}_{i}'^{T}U_{i}\prod_{j\neq i}\mathbf{x}_{-j}$ for every $\mathbf{x}_{i}'\in \Delta_i$. Every finite game admits at least one NE in mixed strategies. The problem of finding an NE can be expressed as:

\vspace{-0.3cm}
\begin{scriptsize}
\begin{align}
v_i - \sum\limits_{a_{-i}\in A_{-i}}U_i(a_i,a_{-i})\cdot \prod_{\underset{j\neq i}{j\in N:}} x_j(a_j)		 &	\geq 0	&	\forall i\in N, a_i\in A_i				\label{nash1}	\\
x_i(a_i)\cdot \Bigg(v_i -\sum\limits_{a_{-i}\in A_{-i}}U_i(a_i,a_{-i})\cdot \nonumber \\
\cdot \prod_{\underset{j\neq i}{j\in N:}} x_j(a_j)	\Bigg)	& = 0		& \forall i\in N, a_i\in A_i	 \label{nash2}	\\
x_i(a_i) & \geq 0	&	\forall i\in N,a_i\in A_i															 \label{nash3}	\\
\sum_{a_i\in A_i} x_i(a_i) & = 1	&	\forall i\in N													 \label{nash4}
\end{align}
\end{scriptsize}
\vspace{-0.3cm}

\noindent Here $v_{i}$ is the expected utility of agent~$i$.  Constraints~(\ref{nash1}) force the expected utility $v_i$ to be non--smaller than the expected utility given by every action $a_i$ available to agent~$i$. Constraints~(\ref{nash2}) force the expected utility $v_i$ of agent~$i$ to be equal to the expected utility given by every action $a_i$ that is played with positive probability by agent~$i$. Constraints~(\ref{nash3}) force each probability $x_i(a_i)$ to be nonnegative. Constraints~(\ref{nash4}) force each probability vector  $\mathbf{x}_i$ to sum to one.

An SNE~\cite{aumann1960} strengthens the NE concept by requiring the strategy profile to be resilient also to multilateral deviations by any coalition of agents. That is, in an SNE no coalition of agents can deviate in a way that strictly increases the expected utility of each member of the coalition, again keeping the strategies of the agents outside the coalition fixed. So, an SNE combines two notions: an SNE is an NE and it is weakly Pareto efficient over the space of all the strategy profiles for each possible coalition. Unlike an NE, an SNE is not assured to exist even in mixed strategies.
%
%
%
%
%
%
%

Multi--objective programming provides Pareto optimality conditions~\cite{multiobjective}, based on Karush--Kuhn--Tucker (KKT) results. These conditions are:

\vspace{-0.35cm}
\begin{scriptsize}
\begin{align}
\sum_i \lambda_i\cdot \nabla f_i(\mathbf{z}) + \sum_j \mu_j\cdot \nabla g_j(\mathbf{z}) + \sum_k \nu_k\cdot \nabla h_k(\mathbf{z}) &= \mathbf{0}		\label{KKT1appendix}	\\
\mu_j\cdot g_j(\mathbf{z}) 	&	= 0		&	\forall j																			 \label{KKT2appendix}	\\
\lambda_i,\mu_j			&	\geq 0 	&	\forall i,j																			 \label{KKT3appendix}	\\
\sum_i\lambda_i = 1																											 \label{KKT4appendix}
\end{align}
\end{scriptsize}
\vspace{-0.35cm}

\noindent where $f_i(\mathbf{z})$ are the objective functions to minimize, $g_j(\mathbf{z})$ are inequality constraints as $g_j(\mathbf{z})\leq 0$, and $h_k$ are equality constraints as $h_k(\mathbf{z})= 0$. The $\lambda_i, \mu_j,\nu_k$ are called KKT multipliers; $\lambda_i$ is the weight of objective function $f_i$, $\mu_j$ is the weight of constraint $g_j$, and $h_k$ is the weight of constraint $h_k$.

The KKT conditions~(\ref{KKT1appendix})--(\ref{KKT4appendix}) are necessary conditions for local Pareto efficiency~\cite{multiobjective}.  For games, we can map these conditions to the case of Pareto efficiency for a single coalition $C\subseteq N$ as follows:
\begin{itemize}
\vspace{-0.1cm}
\item $f_i$ is agent~$i$'s expected utility multiplied by `$-1$' (given that in KKT the objective $f_i$ is to minimize);
\vspace{-0.1cm}
\item $g_j$ is a constraint of the form $-x_i(a_i)\leq 0$;
\vspace{-0.1cm}
\item $h_k$ is a constraint of the form $\sum_{a_i\in A_i}x_i(a_i)-1=0$.
\end{itemize}
\vspace{-0.4cm}

\section{Characterization of mixed--strategy SNEs}
\vspace{-0.1cm}
We study the properties of mixed--strategy SNEs. We focus on the basic case of 2--agent games and we discuss how the reasoning can be extended to games with 3 or more agents.

We denote by $P_{mix}$ and by $P_{cor}$ the sets of points in the agents' utility spaces $\mathbb{E}[U_1]\times\mathbb{E}[U_2]$ that are on the Pareto frontier when the agents play \emph{mixed} and \emph{correlated} strategies, respectively. Obviously, points in $P_{cor}$ non--strictly Pareto dominate points in $P_{mix}$, given that mixed strategies constitute a subset of correlated strategies. In addition, we denote by $P_{mix}(S)$ and $P_{cor}(S)$ the Pareto frontiers in mixed and correlated strategies, respectively, when the game is restricted to the sets of actions in support profile $S$.
\vspace{-0.1cm}
\begin{thm}
Consider a non--degenerate 2--agent game with two actions per agent.  If there is a mixed--strategy SNE, then $P_{mix} = P_{cor}$.
\label{thm1}
\end{thm}
\vspace{-0.1cm}
\emph{Proof}. We can write down the game as follows:
\vspace{-0.25cm}
\[
\begin{small}
\begin{array}{rr|c|c|c|}
\multicolumn{2}{c}{}	&	\multicolumn{2}{c}{\textnormal{agent 2}} \\
\multirow{4}{*}{\begin{sideways}agent 1\end{sideways}}	&		&	\mathsf{a}_3	&	\mathsf{a}_4	\\ \cline{2-4}
	&	\mathsf{a}_1	&	p_{1}, q_{1} 	& 	p_{2}, q_{2}	\\ \cline{2-4}
	&	\mathsf{a}_2	&	p_{3}, q_{3}	&	p_{4}, q_{4}	\\ \cline{2-4}
\end{array}
\end{small}
\]
\vspace{-0.2cm}
By assumption:
\begin{itemize}
\item There is a mixed--strategy NE.  Therefore: 

\vspace{-0.3cm}
\begin{scriptsize}
\begin{align}
x_2(\mathsf{a}_3)\cdot p_1+x_{2}(\mathsf{a}_4)\cdot p_2 & = x_{2}(\mathsf{a}_3)\cdot p_3+x_{2}(\mathsf{a}_4)\cdot p_4	\label{nashthm2-1}	\\
x_1(\mathsf{a}_1)\cdot q_1+x_{1}(\mathsf{a}_2)\cdot q_3 & = x_{1}(\mathsf{a}_1)\cdot q_2+x_{1}(\mathsf{a}_3)\cdot q_4	\label{nashthm2-2}
\end{align}
\end{scriptsize}
\vspace{-0.5cm}

\item The NE is on $P_{mix}$,  being an SNE.   Therefore, the KKT conditions are satisfied (since they are necessary conditions for local weak Pareto efficiency):

\vspace{-0.3cm}
\begin{scriptsize}
\begin{align}
-\lambda_1\cdot (x_2(\mathsf{a}_3)\cdot p_1+x_{2}(\mathsf{a}_4)\cdot p_2)-\qquad\qquad\qquad\qquad\nonumber	\\	-\lambda_2\cdot (x_2(\mathsf{a}_3)\cdot q_1+x_{2}(\mathsf{a}_4)\cdot q_2) & = \nu_1	\label{KKT1} 	\\
-\lambda_1\cdot (x_2(\mathsf{a}_3)\cdot p_3+x_{2}(\mathsf{a}_4)\cdot p_4)-\qquad\qquad\qquad\qquad\nonumber	\\	-\lambda_2\cdot (x_2(\mathsf{a}_3)\cdot q_3+x_{2}(\mathsf{a}_4)\cdot q_4) & = \nu_1	\label{KKT2}	\\
-\lambda_1\cdot (x_1(\mathsf{a}_1)\cdot p_1+x_{1}(\mathsf{a}_3)\cdot p_3)-\qquad\qquad\qquad\qquad\nonumber	\\	-\lambda_2\cdot (x_1(\mathsf{a}_1)\cdot q_1+x_{1}(\mathsf{a}_3)\cdot q_3) & = \nu_2	\label{KKT3}	\\
-\lambda_1\cdot (x_1(\mathsf{a}_1)\cdot p_2+x_{1}(\mathsf{a}_2)\cdot p_4)-\qquad\qquad\qquad\qquad\nonumber	\\	-\lambda_2\cdot (x_1(\mathsf{a}_1)\cdot q_2+x_{1}(\mathsf{a}_3)\cdot q_4) & = \nu_2	\label{KKT4}
\end{align}
\end{scriptsize}
\vspace{-0.3cm}

\end{itemize}
By combining (\ref{nashthm2-1}), (\ref{nashthm2-2}) with (\ref{KKT1})--(\ref{KKT4}), we obtain

\vspace{-0.3cm}
\begin{scriptsize}
\begin{align}
x_2(\mathsf{a}_3)\cdot q_1+x_{2}(\mathsf{a}_4)\cdot q_2 & = x_{2}(\mathsf{a}_3)\cdot q_3+x_{2}(\mathsf{a}_4)\cdot q_4	\label{KKTthm2-1}	\\
x_1(\mathsf{a}_1)\cdot p_1+x_{1}(\mathsf{a}_2)\cdot p_3 & = x_{1}(\mathsf{a}_1)\cdot p_2+x_{1}(\mathsf{a}_2)\cdot p_4	\label{KKTthm2-2}
\end{align}
\end{scriptsize}
\vspace{-0.3cm}

\noindent By trivial mathematics, we can rewrite (\ref{nashthm2-1}), (\ref{nashthm2-2}), (\ref{KKTthm2-1}), (\ref{KKTthm2-2}) as

\vspace{-0.3cm}
\begin{scriptsize}
\begin{align*}
x_2(\mathsf{a}_3)\cdot (p_1-p_3) & = x_{2}(\mathsf{a}_4)\cdot (p_4- p_2)\\
x_1(\mathsf{a}_1)\cdot (q_1-q_2) & = x_{1}(\mathsf{a}_2)\cdot (q_4- q_3)\\
x_2(\mathsf{a}_3)\cdot (q_1-q_3) & = x_{2}(\mathsf{a}_4)\cdot (q_4- q_2)\\
x_1(\mathsf{a}_1)\cdot (p_1-p_2) & = x_{1}(\mathsf{a}_2)\cdot (p_4- p_3)
\end{align*}
\end{scriptsize}
\vspace{-0.3cm}

We can safely assume $p_1\neq p_3$, $p_2\neq p_4$, $p_1 \neq p_2$, $p_3 \neq p_4$, and the analogous inequalities for agent~$2$, since this assumption excludes only degenerate games. Indeed, if $p_1= p_3$, then, by the above conditions, we have $p_2= p_4$, and therefore actions $\mathsf{a}_1$ and $\mathsf{a}_2$ are the same; if $p_1= p_2$, then, by the above conditions, we have $p_3= p_4$, and therefore, in order to have a mixed--strategy NE, we need $p_1= p_2=p_3= p_4$. The same reasoning holds for agent~$2$. Thus, we derive the following conditions:

\vspace{-0.3cm}
\begin{scriptsize}
\begin{align*}
\frac{p_{1}-p_2}{p_4- p_3} & = \frac{q_{1}-q_2}{q_4- q_3}		&	\frac{p_{1}-p_3}{p_4- p_2} & = \frac{q_{1}-q_3}{q_4- q_2}	
\end{align*}
\end{scriptsize}
\vspace{-0.3cm}

\begin{figure}[h]
\begin{minipage}{3.8cm}
\begin{pspicture}*(-1,-0.7)(3.2,2.4)
\scalebox{0.6}{


\psaxes{->}(0,0)(0,0)(5.3,3.6)

\psdots[linecolor=black,dotsize=6pt](1,2)
\psdots[linecolor=black,dotsize=6pt](3,1.5)
\psdots[linecolor=black,dotsize=6pt](2,2.5)
\psdots[linecolor=black,dotsize=6pt](2,1)

\psdots[linecolor=black,dotsize=6pt](2,1.75)

\psline[linecolor=black,linewidth=2pt]{-}(1,2)(2,2.5)(3,1.5)(2,1)(1,2)

\uput{0}[0](0.6,1.6){$R_1$}
\uput{0}[0](3.2,1.7){$R_4$}
\uput{0}[0](2.2,2.7){$R_3$}
\uput{0}[0](1.6,0.6){$R_2$}

\uput{0}[0](1.45,2){NE}

\uput{0}[0](2.2,-1){$\mathbb{E}[U_1]$}
\rput[tr]{90}(-1,2){$\mathbb{E}[U_2]$}
}
\end{pspicture}
\centering
\small(\emph{a})
\end{minipage}
\begin{minipage}{4.2cm}
\begin{pspicture}*(-1,-0.7)(3.2,2.4)
\scalebox{0.6}{


\psaxes{->}(0,0)(0,0)(5.3,3.6)

\psline[linecolor=black,linewidth=2pt]{-}(1,2.5)(4,1)

\psdots[linecolor=black,dotsize=6pt](4,1)
\psdots[linecolor=black,dotsize=6pt](3,1.5)
\psdots[linecolor=black,dotsize=6pt](2,2)
\psdots[linecolor=black,dotsize=6pt](1,2.5)
\psdots[linecolor=black,dotsize=6pt](2.5,1.75)

\uput{0}[0](2,1.4){SNE}

\uput{0}[0](4.2,1.2){$R_1$}
\uput{0}[0](3.2,1.7){$R_4$}
\uput{0}[0](2.2,2.2){$R_3$}
\uput{0}[0](1.2,2.7){$R_2$}

\uput{0}[0](2.2,-1){$\mathbb{E}[U_1]$}
\rput[tr]{90}(-1,2){$\mathbb{E}[U_2]$}
}
\end{pspicture}
\centering
\small(\emph{b})
\end{minipage}
\vspace{-0.2cm}
\caption{Examples used in the proof of Theorem~\ref{thm1}.}
\vspace{-0.2cm}
\label{proofparallelogram}
\end{figure}
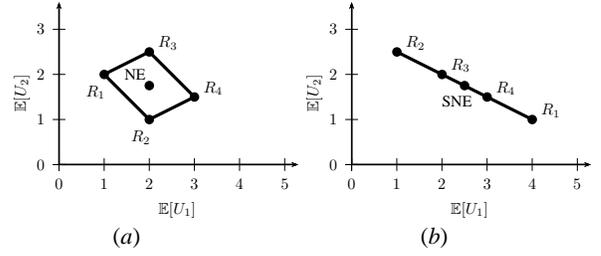

We can give a simple geometric interpretation of the above conditions. Call $R_i = (p_i,q_i)$. Each $R_i$ is a point in the space $\mathbb{E}[U_1]\times \mathbb{E}[U_2]$. The above conditions state that:
\begin{itemize}
\item $\overline{R_1R_2}$ is parallel to $\overline{R_3R_4}$,
\item $\overline{R_1R_3}$ is parallel to $\overline{R_2R_4}$,
\end{itemize}
and therefore $R_1,R_2,R_3,R_4$ are the vertices of a parallelogram, see Fig.~\ref{proofparallelogram}(\emph{a}). Given that
\begin{itemize}
\item a mixed--strategy NE is strictly inside the parallelogram (it being the convex (non--degenerate) combination of the vertices), see Fig.~\ref{proofparallelogram}(\emph{a}), and that
\item it must be on a Pareto efficient edge (since, if it is strictly inside the parallelogram---as in Fig.~\ref{proofparallelogram}(\emph{a})---then it is Pareto dominated by some point on some edge),
\end{itemize}
we have that $R_1,R_2,R_3,R_4$ must be aligned according to a line of the form $\mathbb{E}[U_1]+\phi\cdot\mathbb{E}[U_2]=const$ with $\phi\in(-1,0)$, see, e.g., Fig.~\ref{proofparallelogram}(\emph{b}). Thus, the combination of $R_1,R_2,R_3,R_4$ through every mixed--strategy profile lies on the line connecting the two extreme vertices; for example, in Fig.~\ref{proofparallelogram}(\emph{b}) the extreme vertices are $R_2$ and $R_1$. It trivially follows that $P_{mix}= P_{cor}$.
$\hfill\Box$

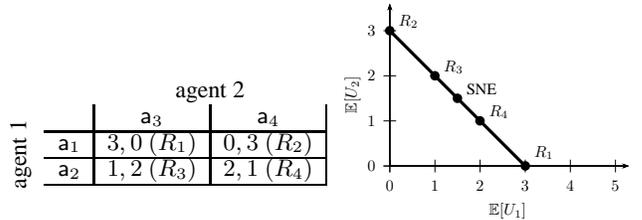
\begin{figure}[h]
\vspace{-0.3cm}
\begin{minipage}{3.9cm}
\[
\begin{small}
\begin{array}{rr|c|c|}
\multicolumn{2}{c}{}	&	\multicolumn{2}{c}{\textnormal{agent 2}} \\
\multirow{4}{*}{\begin{sideways}agent 1\end{sideways}}	&		&	\mathsf{a}_3	&	\mathsf{a}_4	\\ \cline{2-4}
	&	\mathsf{a}_1	&	3,0~(R_1)	& 	0,3~(R_2)	\\ \cline{2-4}
	&	\mathsf{a}_2	&	1,2~(R_3)	&	2,1~(R_4)	\\ \cline{2-4}
\end{array}
\end{small}
\]
\end{minipage}
\begin{minipage}{4.6cm}
\begin{pspicture}*(-1.25,-1)(3.2,2.2)
\scalebox{0.6}{


\psaxes
{->}(0,0)(0,0)(5.3,3.6)

\psline[linecolor=black,linewidth=2pt]{-}(0,3)(3,0)

\psdots[linecolor=black,dotsize=6pt](3,0)
\psdots[linecolor=black,dotsize=6pt](2,1)
\psdots[linecolor=black,dotsize=6pt](1,2)
\psdots[linecolor=black,dotsize=6pt](0,3)
\psdots[linecolor=black,dotsize=6pt](1.5,1.5)

\uput{0}[0](1.7,1.7){SNE}

\uput{0}[0](3.2,0.3){$R_1$}
\uput{0}[0](2.2,1.2){$R_4$}
\uput{0}[0](1.2,2.2){$R_3$}
\uput{0}[0](0.2,3.2){$R_2$}

\uput{0}[0](2.2,-1){$\mathbb{E}[U_1]$}
\rput[tr]{90}(-1,2){$\mathbb{E}[U_2]$}
}
\end{pspicture}
\end{minipage}
\vspace{-0.7cm}
\caption{Example 2--agent game (left) and its Pareto frontier (right).}
\vspace{-0.2cm}
\label{exam:bimatrixSNEalignment2actions}
\end{figure}

The proof of the above theorem provides necessary conditions for a game to admit a mixed--strategy SNE. These conditions are not sufficient. Indeed, we can show that only some alignments of $R_1,R_2,R_3,R_4$ can lead to an SNE:
\vspace{-0.1cm}
\begin{cor}
The only alignments of $R_1,R_2,R_3,R_4$  that can lead to an SNE satisfy the following conditions:
\begin{itemize}
\vspace{-0.1cm}
\item $R_1$ or $R_4$ is one extreme,
\vspace{-0.1cm}
\item moving from the previous extreme, the next vertex is $R_4$ or $R_1$,
\vspace{-0.1cm}
\item the next vertex is $R_2$ or $R_3$,
\vspace{-0.1cm}
\item $R_3$ or $R_2$ is the other extreme.
\vspace{-0.1cm}
\end{itemize}\label{cor:alignments}
\end{cor}
\vspace{-0.1cm}
\emph{Proof}. We initially show that it is not possible to have a mixed--strategy NE for alignments different from those considered in the corollary. For reasons of space, we study a single case, the proof in the other cases is similar. Consider, for contradiction, the case in which $R_1$ is the extreme with the maximum $\mathbb{E}[U_1]$, then the sequence is $R_2$, $R_3$, and $R_4$. 
%
It can be observed that the first action of the first agent dominates the second action because $U_1(R_1)>U_1(R_3)$ and $U_1(R_2)>U_1(R_4)$ and therefore the first agent will play the first action with probability one. So, there is no mixed--strategy NE. Instead, for the alignments considered in the corollary, dominance does not apply, as shown, e.g., in Fig.~\ref{exam:bimatrixSNEalignment2actions}.$\hfill\Box$

\begin{figure}[h]
\vspace{-0.4cm}
\begin{minipage}{4.7cm}
\[
\hspace{-0.2cm}
\begin{scriptsize}
\begin{array}{rr|c|c|c|c|}
\multicolumn{2}{c}{}	&	\multicolumn{4}{c}{\textnormal{agent 2}} \\
	&		&	\mathsf{a}_5	&	\mathsf{a}_6	&	\mathsf{a}_7 	&	\mathsf{a}_8	\\ \cline{2-6}
\multirow{4}{*}{\begin{sideways}agent 1\end{sideways}}	&	\mathsf{a}_1	&	3,0 	& 	0,3	&	\hspace{-0.1cm}-5,-5\hspace{-0.1cm}		 &	\hspace{-0.1cm}-5,-5\hspace{-0.1cm}\\ \cline{2-6}
	&	\mathsf{a}_2	&	1,2	&	2,1	&	\hspace{-0.1cm}-5,-5\hspace{-0.1cm}	&	\hspace{-0.1cm}-5,-5\hspace{-0.1cm}\\ \cline{2-6}
	&	\mathsf{a}_3	&	\hspace{-0.1cm}-5,-5\hspace{-0.1cm}	&	\hspace{-0.1cm}-5,-5\hspace{-0.1cm}	&	5,0	&	0,0\\ \cline{2-6}
	&	\mathsf{a}_4	&	\hspace{-0.1cm}-5,-5\hspace{-0.1cm}	&	\hspace{-0.1cm}-5,-5\hspace{-0.1cm}	&	0,0	&	0,5\\ \cline{2-6}
\end{array}
\end{scriptsize}
\]
\end{minipage}
\begin{minipage}{3.5cm}
\begin{pspicture}*(-1,-1)(3.2,2.7)
\scalebox{0.5}{


\psaxes
{->}(0,0)(0,0)(5.3,5.3)

\psline[linecolor=gray,linestyle=dashed,linewidth=2pt]{-}(3,0)(2.592000,0.392000)

\psline[linecolor=gray,linestyle=dashed,linewidth=2pt]{-}(0,3)(0.392000,2.592000)

\psline[linecolor=gray,linestyle=dashed,linewidth=2pt]{-}(0.392000,2.592000)(0.420500,2.520500)(0.450000,2.450000)(0.480500,2.380500)(0.512000,2.312000)(0.544500,2.244500)(0.578000,2.178000)(0.612500,2.112500)(0.648000,2.048000)(0.684500,1.984500)(0.722000,1.922000)(0.760500,1.860500)(0.800000,1.800000)(0.840500,1.740500)(0.882000,1.682000)(0.924500,1.624500)(0.968000,1.568000)(1.012500,1.512500)(1.058000,1.458000)(1.104500,1.404500)(1.152000,1.352000)(1.200500,1.300500)(1.250000,1.250000)(1.300500,1.200500)(1.352000,1.152000)(1.404500,1.104500)(1.458000,1.058000)(1.512500,1.012500)(1.568000,0.968000)(1.624500,0.924500)(1.682000,0.882000)(1.740500,0.840500)(1.800000,0.800000)(1.860500,0.760500)(1.922000,0.722000)(1.984500,0.684500)(2.048000,0.648000)(2.112500,0.612500)(2.178000,0.578000)(2.244500,0.544500)(2.312000,0.512000)(2.380500,0.480500)(2.450000,0.450000)(2.520500,0.420500)(2.592000,0.392000)(2.664500,0.364500)

\psline[linecolor=black,linewidth=2pt]{-}(0.364500,2.664500)(2.664500,0.364500)

\psline[linecolor=black,linewidth=2pt]{-}(0.000000,5.000000)(0.000500,4.900500)(0.002000,4.802000)(0.004500,4.704500)(0.008000,4.608000)(0.012500,4.512500)(0.018000,4.418000)(0.024500,4.324500)(0.032000,4.232000)(0.040500,4.140500)(0.050000,4.050000)(0.060500,3.960500)(0.072000,3.872000)(0.084500,3.784500)(0.098000,3.698000)(0.112500,3.612500)(0.128000,3.528000)(0.144500,3.444500)(0.162000,3.362000)(0.180500,3.280500)(0.200000,3.200000)(0.220500,3.120500)(0.242000,3.042000)(0.264500,2.964500)(0.288000,2.888000)(0.312500,2.812500)(0.338000,2.738000)(0.364500,2.664500)(0.37,2.66)

\psline[linecolor=black,linewidth=2pt]{-}(2.66,0.37)(2.664500,0.364500)(2.738000,0.338000)(2.812500,0.312500)(2.888000,0.288000)(2.964500,0.264500)(3.042000,0.242000)(3.120500,0.220500)(3.200000,0.200000)(3.280500,0.180500)(3.362000,0.162000)(3.444500,0.144500)(3.528000,0.128000)(3.612500,0.112500)(3.698000,0.098000)(3.784500,0.084500)(3.872000,0.072000)(3.960500,0.060500)(4.050000,0.050000)(4.140500,0.040500)(4.232000,0.032000)(4.324500,0.024500)(4.418000,0.018000)(4.512500,0.012500)(4.608000,0.008000)(4.704500,0.004500)(4.802000,0.002000)(4.900500,0.000500)(5.000000,0.000000)

\psdots[linecolor=black,dotsize=6pt](1.51,1.51)

\uput{0}[0](1.7,1.7){SNE}

\uput{0}[0](2.2,-1){$\mathbb{E}[U_1]$}
\rput[tr]{90}(-1,3){$\mathbb{E}[U_2]$}
}
\end{pspicture}
\end{minipage}
\vspace{-0.7cm}
\caption{Example 2--agent game (left) and its Pareto frontier (right).}
\vspace{-0.2cm}
\label{exam:morethan2actions}
\end{figure}
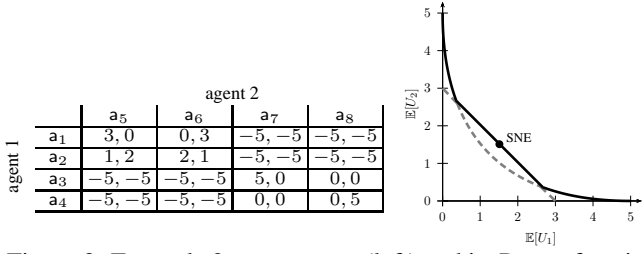

We now extend the previous result to the setting in which each agent has $m$ actions and $|S_1|=|S_2|=2$.
\vspace{-0.1cm}
\begin{cor}
Consider a non--degenerate 2--agent game with $m$ actions per agent.  If there is a mixed--strategy SNE with support sizes $|S_1|=|S_2|=2$, then $P_{mix}(S_1,S_2) 
= P_{cor}(S_1,S_2)$.
\label{cor:supp2mactionsTT}
\end{cor}
\vspace{-0.1cm}
\emph{Proof}. We can split the NE constraints and KKT conditions into two groups: those generated considering deviations towards pure or mixed strategies over the supports $S_1$ and $S_2$ and those generated considering deviations towards pure or mixed strategies over actions off the supports $S_1$ and $S_2$. The constraints belonging to the first group are the same as in the case with two actions per agent considered in the proof of Theorem~\ref{thm1}. The second group overconstrains the problem and it is not necessary for the proof. Thus, restricting the game to the actions in $S_1$ and $S_2$, Theorem~\ref{thm1} holds and therefore $P_{mix}(S_1,S_2) = P_{cor}(S_1,S_2)$, see, e.g., Fig.~\ref{exam:morethan2actions}.$\hfill\Box$

Next we study the case of arbitrary--sized (potentially degenerate) games admitting SNEs with full supports.
\vspace{-0.1cm}
\begin{thm}\label{thm:suppMmactions}
Consider a 2--agent game with $m_1$ and $m_2$. If there is a mixed--strategy SNE with $|S_1|=m_1$ and $|S_2|=m_2$, then $P_{mix} = P_{cor}$.
\end{thm}
\vspace{-0.1cm}
\emph{Proof sketch}. The proof is similar to the proof of Theorem~\ref{thm1}. Define $R_{i,j}=(U_1(i,j),U_2(i,j))$. By NE constraints and KKT conditions, we have that
\begin{itemize}
\vspace{-0.1cm}
\item for all $j$, $\sum_{a_i}x_{1}(a_i)\cdot R_{i,j}=const$  (i.e., the convex combinations of the elements of each column of the bimatrix must have the same value), and
\vspace{-0.1cm}
\item for all $i$, $\sum_{a_j}x_2(a_j)\cdot R_{i,j}=const$ (i.e., the convex combinations of the elements of each column of the bimatrix must have the same value),
\end{itemize}
and we have that
\begin{itemize}
\vspace{-0.1cm}
\item each convex combination is strictly inside the polygon whose vertices are points $R_{i,j}$ (because the combination is not degenerate), and
\vspace{-0.1cm}
\item the combination must be on a Pareto efficient edge (otherwise it would be Pareto dominated),
\end{itemize}
\vspace{-0.1cm}
and therefore all the vertices must be aligned. In this way, the combination of all the points $R_{i,j}$ for every mixed strategy leads to a point that lies on the line connecting all the $R_{i,j}$. Thus we have $P_{mix} = P_{cor}$.$\hfill\Box$

The extension of Corollary~\ref{cor:alignments} to the case with support size larger than two is very involved and we omit it here due to limited space.  Instead, Corollary~\ref{cor:supp2mactionsTT} easily extends to the case in which the support per agent is larger than two (we omit the proof because it is the same as that of  Corollary~\ref{cor:supp2mactionsTT}).
\begin{cor}
Consider a 2--agent game with $m_1$ and $m_2$. If there is a mixed--strategy SNE with $|S_1|=\overline{m}_1$ and $|S_2|=\overline{m}_2$, then $P_{mix}(S_1,S_2) = P_{cor}(S_1,S_2)$.
\end{cor}

We will now discuss how the above results extend to more than two agents. 
For example, in the 3--agent setting, the vector of payoffs for each action profile is $R_{i,j,k}=(U_1(i,j,k),U_2(i,j,k),U_3(i,j,k))$. The crucial result is that necessary conditions, generated for only the actions in the supports, for mixed--strategy SNEs forced by NE constraints with KKT conditions for all the coalitions (i.e., $\{1,2\}$, $\{1,3\}$, $\{2,3\}$, $\{1,2,3\}$)  require that all the $R_{i,j,k}$ lie on a plane (with $n$--agent games, all the payoff vectors restricted on $S$ must lie on an $(n-1)$--dimensional hyperplane). Thus, we have:
\vspace{-0.1cm}
\begin{thm}
Consider an $n$--agent game.  If there is a mixed--strategy SNE with support profile~$S$ then $P_{mix}(S)  = P_{cor}(S)$.
\label{cor:supp2mactions}
\end{thm}
%
%
\vspace{-0.1cm}
The above theorem provides only some necessary conditions, given that many other conditions, due to, e.g., dominance, are required to have mixed--strategy SNEs. Interestingly, we can show that there are vectors of payoffs that satisfy all these conditions. For example, the following 3--agent game has a mixed--strategy SNE in which all the agents randomize with uniform probability over all their actions:

\vspace{-0.4cm}
\[
\begin{scriptsize}
\begin{array}{rr|c|c|c|}
\multicolumn{2}{c}{}	&	\multicolumn{3}{c}{\textnormal{agent 2}} \\
											&				&	\mathsf{a}_4	&	\mathsf{a}_5	&	 \mathsf{a}_6	\\ 	\cline{2-5}
\multirow{3}{*}{\begin{sideways}agent 1\end{sideways}}	&	\mathsf{a}_1	&	1, 0,0 		& 	0, 1,0		 & 	0,0,1			\\ 	\cline{2-5}
											&	\mathsf{a}_2	&	0, 1,0		&	0,0,1			& 	 1,0,0			\\ 	\cline{2-5}
											&	\mathsf{a}_3	&	0,0,1			&	1,0,0			&	 0,1,0			 \\ 	\cline{2-5}
\multicolumn{2}{c}{}& \multicolumn{3}{c}{\mbox{Agent 3 plays action } \mathsf{a}_7}\\											 
\end{array}
\begin{array}{rr|c|c|c|}
\multicolumn{2}{c}{}	&	\multicolumn{3}{c}{\textnormal{agent 2}} \\
											&				&	\mathsf{a}_4	&	\mathsf{a}_5	&	 \mathsf{a}_6	\\ 	\cline{2-5}
\multirow{3}{*}{\begin{sideways}agent 1\end{sideways}}	&	\mathsf{a}_1	&	0,1,0 		& 	0,0,1			 & 	1,0,0			\\ 	\cline{2-5}
											&	\mathsf{a}_2	&	0,0,1			&	1,0,0			& 	 0,1,0			\\ 	\cline{2-5}
											&	\mathsf{a}_3	&	1,0,0			&	0,1,0			&	 0,0,1			 \\ 	\cline{2-5}
\multicolumn{2}{c}{}& \multicolumn{3}{c}{\mbox{Agent 3 plays action } \mathsf{a}_8}\\											 
\end{array}
\end{scriptsize}
\]
\vspace{-0.2cm}
\[
\begin{scriptsize}
\begin{array}{rr|c|c|c|}
\multicolumn{2}{c}{}	&	\multicolumn{3}{c}{\textnormal{agent 2}} \\
											&				&	\mathsf{a}_4	&	\mathsf{a}_5	&	 \mathsf{a}_6	\\ 	\cline{2-5}
\multirow{3}{*}{\begin{sideways}agent 1\end{sideways}}	&	\mathsf{a}_1	&	0, 0,1 		& 	1, 0,0		 & 	0,1,0			\\ 	\cline{2-5}
											&	\mathsf{a}_2	&	1,0,0			&	0,1,0			& 	 0,0,1			\\ 	\cline{2-5}
											&	\mathsf{a}_3	&	0,1,0			&	0,0,1			&	 1,0,0			 \\ 	\cline{2-5}
\multicolumn{2}{c}{}& \multicolumn{3}{c}{\mbox{Agent 3 plays action } \mathsf{a}_9}\\											 
\end{array}
\end{scriptsize}
\]
\vspace{-0.3cm}

\vspace{-0.3cm}
\section{Worst--case 2--agent instances for support--enumeration algorithms}
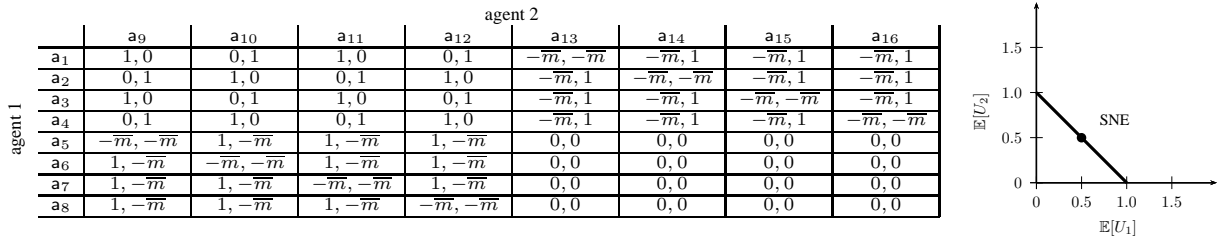
\begin{figure*}
\begin{minipage}{13cm}\[
\begin{scriptsize}
\begin{array}{rr|c|c|c|c|c|c|c|c|}
\multicolumn{2}{c}{}	&	\multicolumn{8}{c}{\textnormal{agent 2}} \\
											&		&	\mathsf{a}_9			&	\mathsf{a}_{10}			 &	\mathsf{a}_{11}		& 	\mathsf{a}_{12}		&	\mathsf{a}_{13}			&	\mathsf{a}_{14}			&	 \mathsf{a}_{15}		& 	\mathsf{a}_{16}			\\ \cline{2-10}
\multirow{8}{*}{\begin{sideways}agent 1\end{sideways}}	&	\mathsf{a}_1	&	1,0		&	0,1		&	 1,0	& 	0,1	&	-\overline{m},-\overline{m}			&	-\overline{m},1			&	-\overline{m},1		& 	 -\overline{m},1			\\ \cline{2-10}
											&	\mathsf{a}_2	&	0,1		&	1,0		&	0,1	& 	1,0	&	 -\overline{m},1			&	-\overline{m},-\overline{m}			&	-\overline{m},1		& 	 -\overline{m},1			\\ \cline{2-10}
											&	\mathsf{a}	_3&	1,0		&	0,1		&	1,0	& 	0,1	&	 -\overline{m},1			&	-\overline{m},1			&	-\overline{m},-\overline{m}		& 	 -\overline{m},1			\\ \cline{2-10}
											&	\mathsf{a}	_4&	0,1		&	1,0		&	0,1	& 	1,0	&	 -\overline{m},1			&	-\overline{m},1			&	-\overline{m},1		& 	 -\overline{m},-\overline{m}			\\ \cline{2-10}\cline{2-10}
											&	\mathsf{a}	_5&	-\overline{m},-\overline{m}	&	 1,-\overline{m}	&	1,-\overline{m}	& 	1,-\overline{m}	&	0,0			&	0,0			&	0,0		& 	 0,0			\\ \cline{2-10}
											&	\mathsf{a}	_6&	1,-\overline{m}	&	 -\overline{m},-\overline{m}	&	1,-\overline{m}	& 	1,-\overline{m}	&	0,0			&	0,0			&	 0,0		& 	0,0			\\ \cline{2-10}
											&	\mathsf{a}	_7&	1,-\overline{m}	&	1,-\overline{m}	&	 -\overline{m},-\overline{m}	& 	1,-\overline{m}	&	0,0			&	0,0			&	0,0		& 	0,0			 \\ \cline{2-10}
											&	\mathsf{a}_8	&	1,-\overline{m}	&	1,-\overline{m}	&	 1,-\overline{m}	& 	-\overline{m},-\overline{m}	&	0,0			&	0,0			&	0,0		& 	0,0			 \\ \cline{2-10}
\end{array}
\end{scriptsize}
\]
\end{minipage}
\begin{minipage}{5cm}
\begin{pspicture}*(-1,-1)(3.2,3.2)
\scalebox{0.6}{

\psset{xunit=2,yunit=2}

\psaxes[Dx=0.5,Dy=0.5]{->}(0,0)(0,0)(2,2)

\psline[linecolor=black,linewidth=2pt]{-}(1,0)(0,1)

\psdots[linecolor=black,dotsize=6pt](.5,.5)

\uput{0}[0](0.7,0.7){SNE}

\uput{0}[0](0.7,-0.5){$\mathbb{E}[U_1]$}
\rput[tr]{90}(-0.7,1){$\mathbb{E}[U_2]$}
}
\end{pspicture}
\end{minipage}
\label{fig:gamewithsupport4}
\vspace{-0.7cm}
\caption{Example game that has an SNE with $|S_1|=|S_2|=4=\overline{m}$ (left), and its Pareto frontier (right).}
\vspace{-0.2cm}
\end{figure*}

We now leverage the results from the previous section to show that it is possible to generate games, with $m$ actions per agent, that have only one SNE, and the support size can be set anywhere in the range $\{1,\ldots,\lceil\frac{m}{2}\rceil\}$. Thus, given $m$, we can generate $O(4^{\frac{m}{2}})$ different game instances, each with a different SNE. These game instances are the worst--case instances for support--enumeration algorithms, given that, for each possible enumeration, it is possible to generate an instance that requires the algorithm to scan an exponential number of supports. 

We first introduce some results that we subsequently exploit to generate our hard instances.

\begin{cor}
Given an even number $\overline{m}$, consider the following two--agent game in which each agent has $m=2\overline{m}$ actions.
\[
U_i = \left[\begin{array}{cc} U_i^{1,1}	&	U_i^{1,2}	\vspace{0.15cm} \\		U_i^{2,1}	&	 U_i^{2,2}\end{array}\right]
\]
where $U_i^{j,k}$ are matrices $\overline{m}\times \overline{m}$ defined as follows:

\vspace{-0.2cm}
\begin{scriptsize}
\begin{align*}
U_1^{1,1}(i,k)&=\begin{cases}1 & \textnormal{if }i+j\textnormal{ odd} \\ 0 & \textnormal{otherwise}\end{cases}			\\
U_1^{2,1}(i,k)&=\begin{cases}-\overline{m} & \textnormal{if }i=j \\ 1 & \textnormal{otherwise}\end{cases}				 \\
U_1^{1,2}(i,k)&=-\overline{m}																		\\
U_1^{2,2}(i,k)&=0																				\\
U_2^{1,1}(i,k)&=\begin{cases}1 & \textnormal{if }i+j\textnormal{ even} \\ 0 & \textnormal{otherwise}\end{cases} \qquad\qquad \qquad \qquad		\\
U_2^{2,1}&=U_1^{1,2}																		\\
U_2^{1,2}&=U_1^{2,1}																		\\
U_2^{2,2}&=U_1^{2,2}						
\end{align*}
\end{scriptsize}
\vspace{-0.2cm}

\noindent This game has an SNE with support sizes $|S_1|=|S_2|=\overline{m}$ and no other SNE.\label{hard-to-solve-even}
\vspace{-0.1cm}
\end{cor}
\emph{Proof}. Denote by $A_i^1$ the first $\overline{m}$ actions of agent~$i$ and by $A_i^2$ the second $\overline{m}$ actions of agent~$i$. For clarity we split the proof into two parts, discussed in the following paragraphs, respectively. An example game is in Fig.~\ref{fig:gamewithsupport4}.

\emph{There is an SNE with $|S_1|=|S_2|=\overline{m}$}. The strategy profile in which each agent~$i$ plays all actions in $A_i^1$ with uniform probability $\frac{1}{\overline{m}}$ is an SNE. All the actions in the supports, i.e., $A_i^1$, provide utility $\frac{1}{2}$, while all the actions off the supports, i.e., $A_i^2$, provide utility $\frac{1\cdot(\overline{m}-1)-\overline{m}}{\overline{m}}=-\frac{1}{\overline{m}}$. Therefore, it is an NE. Furthermore, the outcomes with utilities $(1,0)$ and $(0,1)$ are Pareto efficient, and the others are (weakly) dominated. Thus the Pareto frontier is a line that connects $(1,0)$ to $(0,1)$; therefore the NE is Pareto efficient. So, it is an SNE.

\emph{There is no other SNE}. We focus on the strategy of agent~$2$ (the same reasoning can be applied for agent~$1$). Suppose agent~$2$ adopts a different strategy than the above uniform strategy over $A_2^1$. If agent~$2$ randomizes over a strict subset of $A_2^1$ composed of only odd actions or only even actions, then agent~$1$ has two best responses, one in $A_1^1$ and one in $A_1^2$ (each provides agent~$1$ utility $1$). However, there is no NE of this form, given that, if agent~$1$ randomizes over such best responses, agent~$2$'s best response would be to play some action in $A_2^2$ with a positive utility in place of a utility of $-\overline{m}$ given by the actions in $A_2^1$. If agent~$2$ randomizes over a strict subset of $A_2^1$ with even and odd actions, then agent~$1$ has one or more best responses in $A_1^2$ that provide utility $1$ to agent~$1$. Also in this case, if agent~$1$ randomizes over such best responses, agent~$2$ would play some actions in $A_2^2$ gaining $0$ in place of $-\overline{m}$ given by actions in $A_2^1$. If agent~$2$ randomizes over less than $\overline{m}$ actions in $A_2^1$ and $A_2^2$, then agent~$1$'s best response is to play some action in $A_1^2$. Also in this case agent~$2$'s best response would be to play some action in $A_2^2$. Finally, when both agents play only actions in $A_i^2$ or play more than $\overline{m}$ actions, we can have NEs, but these NEs are Pareto dominated (because they are randomizations over Pareto--dominated outcomes).
$\hfill\Box$

Games with $\overline{m}$ odd are a simple variation w.r.t. the even case (the proof is omitted being very similar):
\vspace{-0.1cm}
\begin{cor}
Given an odd number $\overline{m}$, consider the following two--agent game in which each agent has $m=2\overline{m}$ actions.
\[
U_i = \left[\begin{array}{cc} U_i^{1,1}	&	U_i^{1,2}	\vspace{0.15cm} \\		U_i^{2,1}	&	 U_i^{2,2}\end{array}\right]
\]
where $U_i^{j,k}$ are matrices $\overline{m}\times \overline{m}$ defined as follows:
\begin{itemize}
\vspace{-0.1cm}
\item $U_i^{1,2}$, $U_i^{2,1}$, $U_i^{2,2}$ are generated as in the case of $\overline{m}$ even,
\vspace{-0.1cm}
\item $U_i^{1,1}$ are composed of a submatrix $(\overline{m}-1)\times (\overline{m}-1)$ generated as in the case of $\overline{m}$ even and all the other entries are $\frac{1}{2}$.
\vspace{-0.1cm}
\end{itemize}
This game has an SNE with support sizes $|S_1|=|S_2|=\overline{m}$ and no other SNE.\label{hard-to-solve-odd}
\end{cor}
\vspace{-0.1cm}

We are now ready to define our hard--to--solve instances.
\begin{defn}[Hard--to--solve instances] Given $m$ and $\overline{m}$, if $\overline{m}\in \{2,\ldots,\lfloor\frac{m}{2}\rfloor\}$, a hard--to--solve game instance is composed as follows:
\vspace{-0.1cm}
\begin{itemize}
\item a sub--bimatrix of size $2\overline{m}\times2\overline{m}$ built as described in Corollaries~\ref{hard-to-solve-even} and~\ref{hard-to-solve-odd},
\vspace{-0.1cm}
\item all the other entries are drawn from $\{-\overline{m}\ldots,0\}$ with uniform probability.
\end{itemize}
\vspace{-0.1cm}
If $\overline{m}=1$, all the entries are drawn from $\{-\overline{m}\ldots,0\}$ with uniform probability except for a pure action profile in which both agents have a utility of $1$.
\end{defn}
\begin{thm}
Hard--to--solve instances admit only one SNE.
\end{thm}
\vspace{-0.1cm}
\emph{Proof sketch}. The SNE described in Corollaries~\ref{hard-to-solve-even} and~\ref{hard-to-solve-odd} is still an SNE because all the additional entries are smaller than the expected utility of the SNE. Because these additional entries are Pareto dominated, no additional SNE exists.\hfill$\Box$ 

We leave open the extension to $n$--agent games.
\vspace{-0.3cm}

\section{Smoothed complexity of SNE finding}
Worst--case complexity, being too pessimistic, is often a bad indicator of the actual performance of an algorithm, and average--case complexity is difficult to determine. A newer metric of complexity, called smoothed complexity, has been gaining interest in recent years~\cite{DBLP:journals/corr/abs-1202-1936}. It studies how the introduction of small perturbations affects the worst--case complexity. There might be several models of perturbations. By far the most common perturbation models are the uniform one and the Gaussian one.  

In the case of the SNE-finding problem, given a perturbation $\mathcal{D}_\sigma$ of magnitude $\sigma$, these are defined as follows.
\vspace{-0.1cm}
\begin{itemize}
\item Uniform perturbation: for each agent $i$, every entry in $U_i$ is subjected to an additive perturbation $[-\sigma,+\sigma]$ with uniform probability.
\vspace{-0.1cm}
\item Gaussian perturbation: for each agent $i$, every entry in $U_i$ is subjected to an additive perturbation $[-z,+z]$ with probability $\frac{1}{\sigma\sqrt{2\pi}}e^{-|U_i(j,k)-z|^2/\sigma^2}$.
\end{itemize}
\vspace{-0.1cm}
Denote by $\tilde{U}_i$ the perturbed utility matrix. 

We will first present results for the 2--agent setting.  In the end of this section we show the generalization to any number of agents.

The smoothed running time of an algorithm $\mathcal{A}$ given a perturbation $\mathcal{D}_\sigma$ is defined as

\vspace{-0.2cm}
\begin{scriptsize}
\[
\textsf{smoothed--}t_\mathcal{A}=\mathbb{E}_{\tilde{U}_1,\tilde{U}_2 \sim \mathcal{D}_\sigma}[t_{\mathcal{A}}(\tilde{U}_1,\tilde{U}_2)|U_1,U_2]
\]
\end{scriptsize}
\vspace{-0.3cm}

\noindent where $t_{\mathcal{A}}(\tilde{U}_1,\tilde{U}_2)$ is the running time for the games instance $(\tilde{U}_1,\tilde{U}_2)$. An algorithm has smoothed polynomial time complexity if for all $0<\sigma<1$ there are positive constants $c,k_1,k_2$ such that:

\vspace{-0.2cm}
\begin{scriptsize}
\[
\textsf{smoothed--}t_\mathcal{A}=O(c\cdot m^{k_1}\cdot \sigma^{-k_2})
\]
\end{scriptsize}
\vspace{-0.4cm}

\noindent where $m$ is the size of the game in terms of actions per agent. Basically, a problem is in smoothed--$\mathcal{P}$ if it admits a smoothed polynomial time algorithm. 
\vspace{-0.1cm}
\begin{thm}
Finding an SNE is in smoothed--$\mathcal{P}$.
\end{thm}
\vspace{-0.1cm}
\emph{Proof}. We provide Algorithm~\ref{alg:smoothedSNE}. It has three main parts.

In the first part (Steps~1--3), the algorithm searches for a pure--strategy SNE by enumerating all the pure--strategy profiles and verifying whether each strategy profile is an SNE. The verification is accomplished by checking whether or not NE constraints are satisfied (this can be done in polynomial time in $m$) and by checking whether or not the strategy profile is on the Pareto frontier (this can be done in polynomial time as shown in~\cite{aamasSNE2013}). If an SNE is found, the algorithm returns it. The maximum number of iterations in the first part of the algorithm is $m^2$.

In the second part (Steps~4--7), the algorithm verifies whether there there are strategy profile of support sizes $|S_1|+|S_2|=3$ such that $P_{mix}(S_1,S_2)=P_{cor}(S_1,S_2)$. This can be accomplished in time $O(m^3)$ by checking whether there is a line connecting at least three entries of all the sub--bimatrices of size $2\times2$. In the affirmative case, the temporary variable \textsf{temp} is set \textsf{true}. Otherwise, \textsf{temp} is set \textsf{false}.

In the third part (Steps~8--13), if \textsf{temp} is \textsf{false}, the algorithm returns \textsf{non--existence}, given that, by Theorem~\ref{thm:suppMmactions}, there is no mixed--strategy SNE. Otherwise, the algorithm enumerates all supports to find mixed--strategy NEs, and for each of them the algorithm verifies whether it is an SNE as done in Steps~1--3. 
In the latter case, the algorithm can take exponential time as discussed in the previous section.

Thus, the running time of Algorithm~\ref{alg:smoothedSNE} is super--polynomial only if it needs to enumerate supports during Steps~8--13 (this can take exponential time). This happens only when $P_{mix}(S_1,S_2)=P_{cor}(S_1,S_2)$ for some $S_1,S_2$ with $|S_1|+|S_2|=3$. Given that the perturbations $\mathcal{D}_\sigma$ over all the entries of the utility matrices are independent and identically distributed, the probability that the perturbed entries are aligned as required by Corollary~3.5 to have SNEs with $|S_1|=|S_2|>1$ is zero. Therefore, the smoothed running time of Algorithm~\ref{alg:smoothedSNE} is polynomial in $m$ and independent of $\sigma$ (for both uniform and Gaussian perturbations). 
$\hfill\Box$

The above result shows that, except for a space of the parameters with zero measure, games admit only pure--strategy SNEs and therefore that verifying the existence of an SNE and finding them is  computationally easy. Interestingly, the instability is due to the combination of NE constraints and Pareto efficiency constraints. Indeed, both the problem of finding an NE and the problem of finding Pareto efficient strategies are not sensitive to perturbations. Thus, while an approximate NE can be found by perturbing a game and finding an NE of the perturbed game (exactly, an NE of an $\epsilon$--perturbed game is an $2\epsilon$--NE of the original game), this is not the case with SNE. Indeed, if a game admits only mixed--strategy SNEs, once perturbed, it does not admit any SNE.

\begin{algorithm}
\begin{algorithmic}[1]
\begin{small}
\FORALL{pure--strategy support profiles}
	\IF{the support profile has an SNE}
		\RETURN the SNE
	\ENDIF
\ENDFOR
\STATE $\mathsf{temp}\leftarrow \mathsf{false}$
\FORALL{strategy profiles with $|S_1|+|S_2|=3$}
	\IF{payoffs restricted to supports are aligned}
		\STATE $\mathsf{temp}\leftarrow \mathsf{true}$
	\ENDIF
\ENDFOR
\IF {not $\mathsf{temp}$}
	\RETURN \textsf{non--existence}
\ELSE
\FORALL{support profiles}
	\IF{the support profile has an SNE}
		\RETURN the SNE
	\ENDIF
\ENDFOR
\ENDIF
\end{small}
\end{algorithmic}
\caption{$\mathsf{SNEfinding}$}
\label{alg:smoothedSNE}
\end{algorithm}

The above results extend to the setting with any (constant) number of agents. As discussed in Section~3, in order to have mixed--strategy SNEs with $n$ agents, all the payoffs vectors of the actions in the agents' supports must lie on the same $(n-1)$--hyperplane. Thus, if perturbed even with small perturbations, the probability that the payoff vectors satisfy such constraints is zero.  Therefore, generic game instances (i.e., all instances except knife--edge cases) have only pure--strategy SNEs (if any).
\vspace{-0.2cm}

\section{Conclusions and future research}
Strong Nash equilibrium (SNE) is the most natural solution concept for games where agents can form coalitions, i.e., coordinate their strategies.  An SNE is a strategy profile where no coalition can deviate in a way that every one of the deviating agents (strictly) benefits.  Given a finite game, a Nash equilibrium always exists, but an SNE might not exist.  For finding an SNE, there has been a shortage of algorithms.  Most algorithms only find pure--strategy SNEs in special game classes.  A recent algorithm finds SNEs generally, but only in 2--agent games.

Our central result is that, in order for a game to have at least one mixed--strategy (i.e., non--pure--strategy) SNE, the agents' payoffs restricted to the agents' supports must, in the case of two agents, lie on the same line, and, in the case of $n$ agents, lie on an $(n-1)$--dimensional hyperplane. Leveraging this result, we provided two contributions. First, we developed worst--case game instances for support--enumeration algorithms. These instances have only one SNE and the support size can be chosen to be of any size---in particular, arbitrarily large. In this way, for each possible enumeration, it is possible to generate an instance that requires the algorithm to scan an exponential number of supports. Second, we proved that, unlike Nash equilibrium, finding an SNE is in smoothed polynomial time: generic game instances (i.e., all instances except knife--edge cases) have only pure--strategy SNEs.

In future research we plan to study the computational complexity of approximating SNE and to design algorithms to do so.  We also plan to study computational issues related to \emph{strong correlated equilibrium}. This concept should present different properties than SNE, e.g., the convexity of the Pareto frontier with this solution concept could make the computation of an equilibrium easier and could make equilibria not sensitive to small perturbations.

\newpage
\bibliography{bib}
\bibliographystyle{named}

\end{document}